# Non-Hermitian Mode Cleaning in Periodically Modulated Multimode Fibers

Mohammad Nayeem Akhter[1], Salim B. Ivars[1], Muriel Botey[1], Ramon Herrero[1] and Kestutis Staliunas[1,2,3]

[1]Universitat Politecnica Catalunya, Dep. de Fisica, Rambla Sant Nebridi 22, 08222, Terrassa (Barcelona) Spain
[2]ICREA, Passeig Lluís Companys 23, 08010, Barcelona, Spain
[3]Vilnius University, Faculty of Physics, Laser Research Center, Sauletekio Ave. 10, 10223, Vilnius, Lithuania

*Corresponding author: mohammad.nayeem.akhter@upc.edu

**Abstract:** We show that the simultaneous modulation of the propagation constant and of the gain/loss coefficient along the multimode fibers results in unidirectional coupling among the modes, which, depending on the modulation parameters, leads to the enhancement or reduction of the excitation of higher order transverse modes. In the latter case, effective mode-cleaning is predicted, in ideal case resulting in single-mode spatially coherent output. The effect is semi-analytically predicted on a simplified Gaussian beam approximation and numerically proven by solving the wave propagation equation introducing the modulated potential.

**Introduction:** Multimode fibers (MMFs) are currently boosting renewed attention, yet generally showing a random (speckle) output, even when high spatial quality beams are injected. This is due to the different propagation constants of the fiber modes. Thus, a coherent ensemble of modes in the input radiation, dephases in propagation along the fiber. Such randomization occurs even for GRaded INdex (GRIN) multimode fibers with parabolic index profile, although the mode propagation constants are equidistant, and in principle such dephasing should not occur (periodic self-imaging is expected instead). However, the smallest imperfections along GRIN fibers (e.g., bending, or stretching) break the mode equidistance, and a random structure of the output field is generally observed. Moreover, in high intensity regimes, the additional randomization appears due to nonlinearities, typically the Kerr or Raman nonlinearities.

Attempts to overcome such dephasing and randomization problems have been reported, for instance, by adaptively adjusting of the phases of the modes of the input beam [1], among other means [2-4]. A recently uncovered effect of the decrease of randomness, the so-called, beam self-cleaning, was proposed in Ref. [5], and further explored [6, 7]. We note, however, that the self-cleaning preserves the integral quantities of the beam, like second order momenta in the direct and wavenumber space, redistributing the energy among the modes. Therefore, it does not lead to a direct reduction of the beam quality parameter, $M^2$.

Generally, the reduction of turbulence in optical fibers cannot be achieved by conventional means. For instance, the modulation of the fiber parameters along the propagation direction (modulation of refraction index, dispersion, nonlinearity) can indeed lead to mode coupling effects in the linear case [8,9], or to the parametric (Faraday) modulational instabilities in nonlinear case [10-12], which generally broaden the angular spectra, but not to a reduction of the turbulence, neither to mode-cleaning of the spatial structure of the beam.

The situation may be substantially different using a periodic **non-Hermitian** potential modulation along the fiber, when not only the Hermitian part (the refraction index, equivalently the propagation constant) is modulated, but also the non-Hermitian part (the gain/loss) is structured. The field of open dissipative non-Hermitian physics gained its popularity from the recently proposed Parity-Time (PT) symmetric systems in the frame of Quantum Mechanics that soon found realizations in optics [13-15]. Yet, PT-symmetric systems represent a small sub-manifold of the periodic non-Hermitian systems, where the phase shift between real and imaginary parts of the potential is restricted to particular values, namely $\pi/2$ or $3\pi/2$ (for anti-PT-symmetry). Also, the application of dynamical non-Hermitian potentials to control the turbulence in some model systems has been proposed [16-18].

We here consider a harmonic non-Hermitian modulation of the propagation constant and the gain along a GRIN MMF with parabolic refraction index profile in transverse direction, as schematically illustrated in Fig.1a. The modulation of the propagation constant may be induced by a periodic variation of the refractive index, for instance, by doping the fiber core or modulating the fiber core radius [19, 20]. It is well known that such a modulation causes a **symmetric** coupling between the modes, as schematically indicated by the blue arrows in Fig.1c. In turn, we assume a modulation of the gain/loss profile of the MMF, with a particular spatial delay with respect to the Hermitian part of the modulation. The main idea is that the simultaneous action of both, the Hermitian and non-Hermitian parts of the modulation along the fiber may eventually induce a **unidirectional** coupling between the modes, as represented by the green arrows in Fig.1c. The unidirectional character of such coupling, and therefore the corresponding energy flow, may be either directed toward higher or lower index modes, depending on the shift between both modulations. For particular phase shifts, this is expected to eventually suppress the higher order modes, resulting in a lowest-mode coherent output. This mode-cleaning mechanism is the basic aim of the present letter.

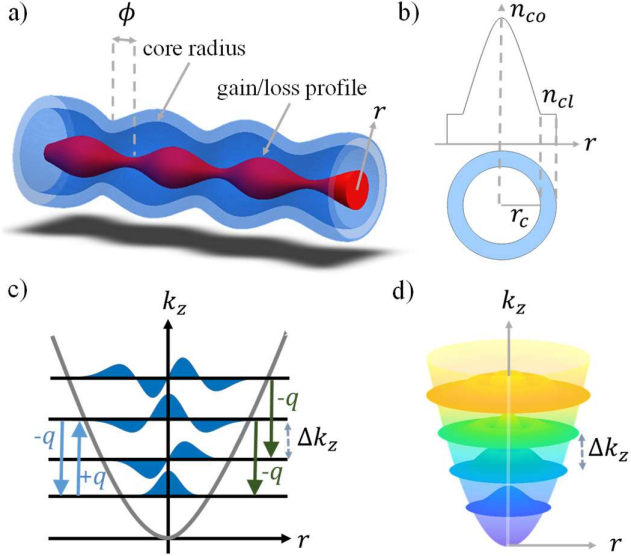

*FIG.1. Modulated GRIN MMF. (a) Periodic modulation on the core radius (Hermitian part) and gain profile (non-Hermitian part). The spatial displacement between the modulations of the fiber core radius and of gain is indicated by a phase shift $\phi$. (b) GRIN fiber cross-section with a parabolic variation of the index. c) Spatial profile and cross-section of the first equidistant eigenmodes. d) 3D mode profiles.*

Here we substantiate this idea: we show that the non-Hermitian modulation of the potential along the fiber indeed results in a tunable distribution of modes at the output. We first derive and explore a simplified model based on a Gaussian beam approximation, which predicts the effect, uncovering analytic insights, and provides estimations of the parameters. The proposal is then proven by direct numerical integration of the wave propagation equation along the fiber. On both models we analyze the modal energy distributions and, as the main result, we demonstrate a substantial condensation of radiation into the lowest order mode; resulting in a ***non-Hermitian mode-cleaning***.

**Full Model:** The propagation of light beams in MMFs with a parabolic refraction index profile may be described by a linear Schrodinger equation as:

$$\frac{\partial A}{\partial z} = i\frac{1}{2}\nabla^2 A - i\frac{\Delta}{r_c^2}r^2 A + i\,V(r,z)A \qquad (1)$$

where $A(x, y, z)$ is the complex field amplitude envelope in the paraxial approximation evolving along $z$. The space coordinates are normalized to $k_0^{-1} = \lambda/2\pi$; where $k_0 = \omega_0 n_{co}/c$ is the light wavenumber, $\nabla^2 = \partial^2/\partial x^2 + \partial^2/\partial y^2$ is the Laplacian in transverse space, $r_c$ is the core radius, $\Delta = (n_{co}^2 - n_{cl}^2)/(2n_{co}^2)$ is the relative index difference, and $n_{co}$ ($n_{cl}$) is the refractive index of the fiber core (cladding), respectively. We neglect frequency dispersion effects (as either continuous wave or sufficiently long pulses are considered), nor the nonlinear effects, along with the Raman scattering.

In the absence of the potential modulation, $V(r,z) = 0$, the fields propagating in MMF exhibit a periodic self-imaging, due to equidistant mode propagation constants. The mode spacing ($\Delta k_z$) and the self-imaging period ($\zeta$) are, respectively, given by:

$$\Delta k_z = \frac{\sqrt{2\Delta}}{r_c}, \quad \zeta = \frac{2\pi}{\Delta k_z} = \frac{\pi r_c}{\sqrt{2\Delta}} \qquad (2)$$

as follows, from Eq. (1). In the presence of a periodic potential with a periodicity close to multiples of the self-imaging period, the modes become resonantly coupled. Note that a transversally uniform potential modulation does not cause any spatial effect *i.e.* any effective coupling between transverse modes, and can therefore be eliminated from Eq. (1) by renormalization of the amplitude and phase of the propagating beam. We therefore assume a periodic modulation in $z$ with a complex transverse profile, in $r$: $V_{Re/Im}(r,z) = V_{Re/Im}(z)e^{-r^2/r_0^2}$. Note that the refractive index and gain modulation may generally present different profiles, however this does not lead to essential differences. The fundamental concept behind this work is to affect the mode distribution by breaking the symmetric mode coupling. Such symmetry breaking may be achieved under the introduction of a longitudinal non-Hermitian modulation in the fiber, to couple modes in a unidirectional way. Thus, we assume a complex harmonic potential in the general form:

$$V(r,z) = [m_1 cos(qz) + im_2 cos(qz + \phi)]e^{-r^2/r_0^2} \qquad (3)$$

where $m_1$ and $m_2$ are the amplitudes of the refractive index and gain/loss modulations respectively and $\phi$ is the delay in the longitudinal direction between these two modulations.

**Gaussian Ansatz:** As a simple approximation of the system dynamics near the lowest transverse mode *i.e.* the lowest order Laguerre-Gauss mode $LG_{00}$, we assume an oscillatory Gaussian ansatz of the form:

$$A(r,z) = \sqrt{\rho(z)}\,e^{-\beta(z)r^2} \qquad (4)$$

with real-valued beam amplitude $\rho(z)$, and complex-valued beam waist $\beta(z) = \beta_r(z) + i\beta_i(z)$. The evolution of these variables, derived from Eq. (1) within the parabolic approximation, is given by:

$$\frac{d\beta_r}{dz} = 4\beta_i\beta_r - \frac{m_2}{r_0^2}cos(qz + \phi) \qquad (5.a)$$

$$\frac{d\beta_i}{dz} = 2(\beta_i^2 - \beta_r^2) + b + \frac{m_1}{r_0^2}cos(qz) \qquad (5.b)$$

$$\frac{d\rho}{dz} = 4\beta_i\rho - 2m_2 cos(qz + \phi)\,\rho \qquad (5.c)$$

where $b = \Delta/r_c^2$.

In absence of modulation, $m_1 = m_2 = 0$, the stationary solution of the above system corresponds to the excitation of the single lowest mode of the fiber ($\beta_{r0} = \sqrt{b/2}$, $\beta_{i0} = 0$ and $\rho_0$). This regime can be achieved by a particular spatial shape of the injection into the fiber (Gaussian beam matching the width of the lowest mode). On the contrary, a mismatched injection ($\beta_{r0} \neq \sqrt{b/2}$, $\beta_{i0} \neq 0$) results in periodic solutions, with the resonance wavenumber $q = q_{res} = 2\pi/\zeta = (2\sqrt{2\Delta})/r_c$, as it follows from Eq. (2), which correspond to a multimode excitation of the autonomous system. Periodic solutions with low amplitude harmonic oscillations can be found by linearizing the system of Eq. (5) around its stationary solution. Such oscillations correspond to the excitation of the fundamental Gaussian mode together with higher order modes of low amplitude. The larger is the modulation amplitude of the periodic solution, the stronger is the excitation of higher modes. Nonharmonic periodic solutions of Eq. (5) correspond to a set of higher order modes.

The **non-autonomous** case ($m_1, m_2 \neq 0$) is more involved as it introduces the driving frequency, $q$, and additional stability conditions for the periodic solutions. In this case, we may rewrite Eq. (5) in the more compact vectorial form:

$$\frac{d\vec{f}}{dz} = \widehat{NL}\vec{f} + \vec{p}\, e^{iqz} \qquad (6)$$

where $\widehat{NL}$ is the nonlinear evolution operator of the autonomous part of (5) acting on the state vector $\vec{f} = (\beta_r, \beta_i, \rho)$, and $\vec{p} = (p_1, p_2, p_3)$ is the vector of the driving amplitudes:

$$p_1 = -\frac{m_2}{r_0^2}e^{i\phi}, p_2 = \frac{m_1}{r_0^2} \text{ and } p_3 = -2m_2 e^{i\phi} \qquad (7)$$

The solution of the driven system can be expressed as the stationary solution of the autonomous system with additional perturbations from the driving:

$$\vec{f}(z) = \vec{f_0} + \vec{\Delta f}\, e^{iqz} + c.c. \qquad (8)$$

where $\vec{f_0} = (\sqrt{b/2}, 0, \rho_0)$ is the stationary state vector, and $\vec{\Delta f}$ is the vector of driven oscillation amplitudes.

Linearization of Eq. (6) with respect to small amplitude driven oscillations leads to:

$$\hat{L} * \vec{\Delta f} = \vec{p} \qquad (9)$$

where $\hat{L}$ is the Jacobian of autonomous system of Eq. (5). The solution of Eq. (9) is obtained by inverting the Jacobian: $\vec{\Delta f} = \widehat{L^{-1}} * \vec{p}$, and reads explicitly:

$$\vec{\Delta f} = \left(\frac{iq\,p_1 + q_{res}p_2}{q_{res}^2 - q^2}, \frac{iqp_2 - q_{res}p_1}{q_{res}^2 - q^2}, \frac{iq_{res}^2 p_1 \rho_0}{q\beta_{r0}(q_{res}^2 - q^2)} + \frac{q_{res}\rho_0 p_2}{\beta_{r0}(q_{res}^2 - q^2)} - \frac{ip_3\rho_0}{q}\right) \qquad (10)$$

where $q_{res} = 8d\beta_{r0}$.

The final expression of the driven oscillations of the beam waist parameter reads:

$$\Delta\beta_r = \frac{\sqrt{q^2 m_2^2 + q_{res}m_1(q_{res}m_1 + 2m_2 q \sin(\phi))}}{r_0^2(q_{res}^2 - q^2)} \qquad (11)$$

which evidences the resonant character of the solution.

**Results:** We calculate numerically these stationary variables, and plot them in parameter space. We also numerically integrate the system of equations (5) to obtain some insights into the transient dynamics, and in order to check the stability of the solutions obtained by the multiscale analysis, as above described.

The parameter space to explore is quite large. After normalization, Eq. (1) contains only one free parameter, $\Delta/r_c^2$, for unmodulated fiber, however the modulation involves five relevant parameters: the real and imaginary parts of the amplitudes of the modulation $m_1$ and $m_2$, their mutual shift $\phi$, the perturbation wavenumber $q$, and the radius of the spatial profile $r_0$ of the modulation of potential. Therefore, we could explore only the most relevant part of the parameter space. For convenience, we introduce $m$ and $\theta$ as: $m_1 = m\cos(\theta), m_2 = m\sin(\theta)$, and numerically explore the parameter space $(\phi, \theta)$ for a small constant value of $m$.

We first analyze the modulation frequency below the resonance frequency ($q < q_{res}$). Figure 2a. summarizes the observation. The oscillations amplitude is maximum or minimum for $\phi = \pi/2$ or $\phi = 3\pi/2$. For a particular ratio between the quadratures $m_1/m_2$ (i.e. for particular $\theta$), the oscillations can vanish to zero. This occurs at the shift $\phi = 3\pi/2$. Actually, Eq. (11) interprets the situation, which tells us that $\Delta\beta_r$ is minimum for $sin(\phi) = -1$ and is maximum for $sin(\phi) = 1$. So we can write in general:

$$\left|\frac{q\,m_2 - q_{res}m_1}{r_0^2(q_{res}^2 - q^2)}\right| < \Delta\beta_r < \left|\frac{q\,m_2 + q_{res}m_1}{r_0^2(q_{res}^2 - q^2)}\right| \qquad (12)$$

Fig. 2b shows the variation of the oscillation amplitude with $\phi$ for $\theta = \pi/4$, corresponding to the white dashed line on the Fig. 2a. The blue circular dots correspond to the numerical values calculated from the full model described by Eq. (1), showing a good agreement between the simplified and full models.

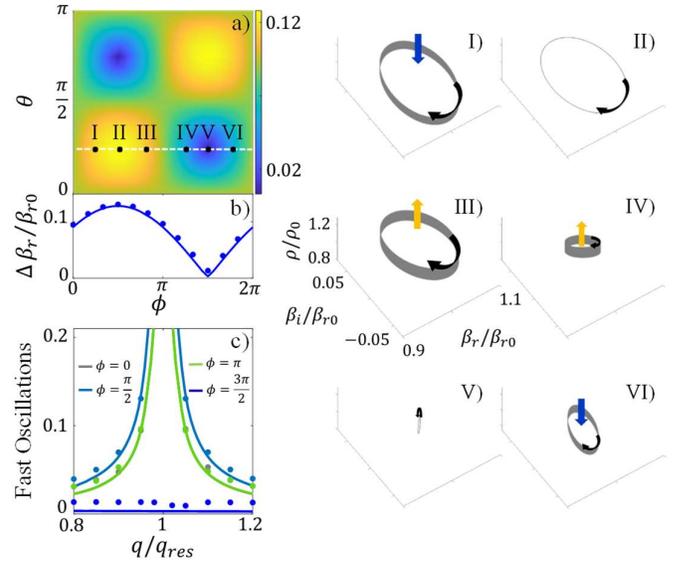

**FIG.2.** *a) Map of the normalized amplitude of the beam width oscillations $\Delta\beta_r/\beta_{r0}$, in the $(\phi, \theta)$ parameter space. b) Cross-section of the map a) along the white dashed line. The solid line in b) is the result from numerical integration of the simplified model, Eq (5), superimposed to Eq. (11). The circular dots obtained by the full model, Eq (1). The six plots on the right hand side show the evolution of the beam in the phase space $(\beta_r/\beta_{r0}, \beta_i/\beta_{r0}, \rho/\rho_0)$, corresponding to different phase shifts between the two modulations, $\phi$, as labeled on the map a) with roman numerals, for: $q = 0.95\, q_{res}, m = 8 \times 10^{-5}$. (c) Dependency of the fast oscillation on modulation frequency q for $\phi = 0, \pi/2, \pi, 3\pi/2$. The lines correspond to the integration of the simplified*

*Gaussian model, Eq. (5) while the dots are values obtained by the full model, Eq. (1). Lines $\phi = 0$ and $\phi = \pi$ are superposed.*

The six plots on the right-hand side of Fig. 2 depict the evolution of the system, Eq. (5), in its phase-space, and correspond to six different situations, fibers with different applied non-Hermitian potentials, leading to different regimes.

We also explore the dependency of the driven oscillations amplitude, as a function of the modulation frequency of the potential. We observe that the oscillation amplitude increases as the modulation frequency approaches the resonance frequency ($q_{res}$) for the cases: $\phi = 0, \pi/2$ and $\pi$ as shown in Figs. 2c. The effect is maximal at $\phi = \pi/2$ yet no effect is observed for $\phi = 3\pi/2$. We observe that the effect is same for phases $\phi = 0$ and $\pi$. We also observe that the effect is symmetric as the modulation frequency passes the resonance frequency.

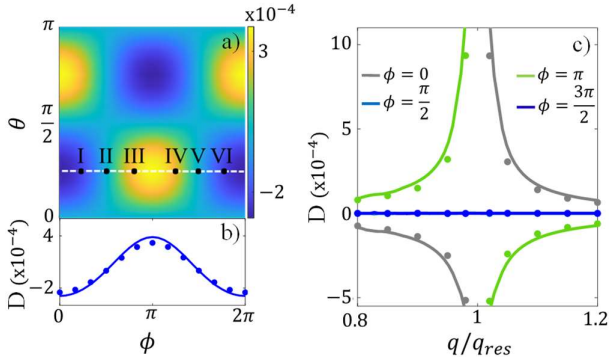

**FIG.3.** *a) Map of the normalized slow drift of the intensity, $D = (\zeta\, d\rho/dz)/\rho_0$, in the $(\phi, \theta)$ parameter space. b) Cross-section of the map a) along the white dashed line. (c) Dependency of the slow drift on the modulation frequency q, for different values of $\phi$, $0$, $\pi/2, \pi, 3\pi/2$. In b) and c) the lines correspond to the integration of the simplified Gaussian model, Eq. (5), while the dots are values obtained from the full model, Eq (1).*

Interestingly, on top of the driven oscillations with the amplitude depending on the position in the phase space of non-Hermitian driving force, the trajectory slowly drifts in the phase space. The arrows in the six plots on the right hand side in Fig.2 show this behaviour. This means that the average photon number increases or decreases in propagation along the fiber. Figure 3a shows the dependency of the slow drift on $(\theta, \phi)$ space. We observe that the sign of the intensity drift is controlled by the both parameters, $\phi$ and $\theta$. For $m_1 = m_2$ ($\theta = \pi/4$), the drift has a maximal dependence on $\phi$ showing a minimum for $\phi = 0$ and a maximum for $\phi = \pi$, see Fig. 3a. Figure 3b shows the variation of the slow drift with $\phi$ for $\theta = \pi/4$, corresponding to the white dashed line on the Fig. 3a. The blue circular dots correspond to the numerical values calculated from the full model described by Eq. (1), showing a good agreement between the simplified and full models. Cases with a negative drift (Fig 2. I and VI) lead to a stable cycle in the $(\beta_r, \beta_i)$ plane, while a positive drift (Fig 2. III and IV) corresponds to unstable cycles, meaning that the beam tends to leave the Gaussian profile. Keeping the same modulation amplitude, this attainment of the limit cycle in the $(\beta_r, \beta_i)$ plane is faster for $\phi = 0$, and requires longer propagation distances as

$\phi$ goes from 0 to $\pi/2$. Figures 3c shows the dependency of the slow drift, as a function of the potential modulation frequency. We observe that for phases $\phi = 0$ and $\phi = \pi$ there is an asymptotic behavior near the resonance frequency ($q_{res}$). Also, we observe that the intensity drift inverts its sign as the modulation frequency passes the resonance frequency and that the effect is reversed for $\phi = \pi$ from $\phi = 0$. We observe no such effect for the no drift cases corresponding to the phases $\phi = \pi/2$ and $\phi = 3\pi/2$ as shown in Figs. 3c.

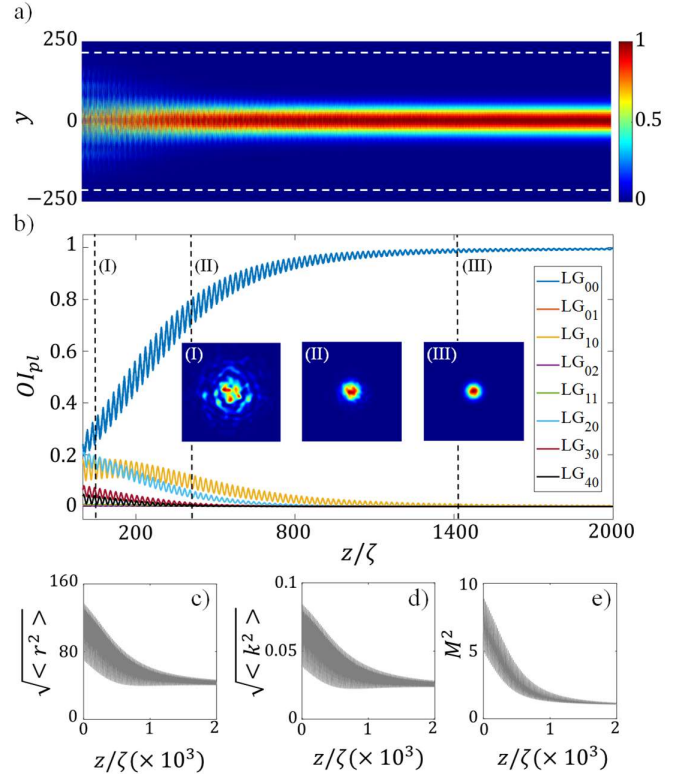

**FIG.4.** *Evolution of the noisy beam along the modulated GRIN MMF. a) Evolution of the transverse noisy profile along the fiber. The white dashed lines indicate the core radius. b) Mode participation of some low order modes along the fiber. Insets: Two-dimensional transverse distributions at different propagation lengths, namely; (I) $z = 50\zeta$, (II) $z = 420\zeta$ and (III) $z = 1410\zeta$. c) Evolution of the beam width, $\sqrt{<r^2>}$; d) angular width, $\sqrt{<k^2>}$; and e) beam quality factor, $M^2$, along the propagation distance. Fiber length is normalized to the self-imaging period $\zeta$, and $m_1 = m_2 = 2 \times 10^{-4}, \phi = 5.5, q = 0.95\, q_{res}$, and $\gamma_0 = 4 \times 10^{-7}$.*

**Full numerical integration:** The previous results predict the control of the oscillation amplitudes, demonstrating they can be reduced, leading to a monomode profile, for optimal parameters. Therefore, predicting the possibility of spatial mode-cleaning of the nosy beam. Next, we further demonstrate the effect by numerical integration of the full model of Eq. (1). A small homogeneous gain term, namely $+\gamma_0 A$, is added on the right-hand side of Eq. (1) to compensate for intensity losses, to conserve the total number of the photons. Speaking in experimental terms – this may be achieved by applying a slightly disbalanced gain modulation. An example of the integration, showing the noisy beam evolution along the fiber is displayed in Fig. 4. Note that the highly multimodal input distribution of the beam inside the MMF is gradually 'attracted' towards a bell-shaped transverse profile, as propagating along the fiber as shown in Fig. 4a. The insets of the Fig. 4b show the two dimensional transverse distribution of the beam at different propagation lengths. Figure 4b depicts the relative participation ($OI_{pl}$) of some Laguerre-Gauss modes, $LG_{pl}$, with low mode order $N$ ($N = 2p + |l| + 1$), in the total field along the fiber. The participation of the lowest order mode ($LG_{00}$) increases tending to 1 as the beam propagates along the fiber, while the participation of higher order modes decrease. Further to characterize the beam cleaning, we calculate the evolution of the beam width in real space, angular width in Fourier space (divergence), and the beam quality factor $M^2$. Figures Figs 4c-4e show a significant reduction of the beam width in both direct and wavenumber space, as the beam quality factor $M^2$ rapidly approaches unity, acquiring an almost Gaussian beam profile.

**Conclusions:** In conclusion, we propose and demonstrate a powerful mechanism for an effective spatial mode cleaning in GRIN MMFs. The proposal is based on the asymmetric mode coupling induced by the introduction of a non-Hermitian modulation of propagation constant and of the gain/loss coefficient along. The fiber may be modeled by a (2+1) D Schrödinger equation with a non-Hermitian potential, where the control over the coupling among transverse modes is mainly governed by the spatial shift between the real and imaginary parts of the complex potential. The effect is first semi-analytically predicted on a simplified Gaussian beam approximation. This model provides a physical insight on the proposal, since it allows estimating the different regimes of unidirectional mode coupling either leading to spatial mode cleaning or to higher mode excitation. The results of the integration on the full model, provides a clear numerical proof of the proposal showing a significant mode-cleaning, irrespectively of the initial intensity. The demonstrated scheme could be experimentally realized within the current nanofabrication technologies, by modulating the core radius of an amplifying fiber of length on the order of meters with distributed losses. Finally, we note that the case of coupling towards high order modes may enhance pulsing and eventually help in super-continuum generation.

**Acknowledgments:** This work has received funding from Horizon 2020 program project MEFISTA (Project No 861152), from Spanish Ministry of Science, Innovation and Universities (MICINN) under grant PID2019-109175GB-C2, and from European Social Fund (Project No 09.3.3- LMT-K712-17- 0016) under grant agreement with the Research Council of Lithuania (LMTLT).